\def \SAIT #1 #2 {{\em Mem.\ Soc.\ Astron.\ It.\/} {\bf #1}, #2}
\def \MESS #1 #2 {{\em The Messenger\/} {\bf #1}, #2}
\def \ASTRNACH #1 #2 {{\em Astron. Nach.\/} {\bf #1}, #2}
\def \AAP #1 #2 {{\em Astron. Astrophys.\/} {\bf #1}, #2}
\def \AAL #1 #2 {{\em Astron. Astrophys. Lett.\/} {\bf #1}, L#2}
\def \AAR #1 #2 {{\em Astron. Astrophys. Rev.\/} {\bf #1}, #2}
\def \AAS #1 #2 {{\em Astron. Astrophys. Suppl. Ser.\/} {\bf #1}, #2}
\def \AJ #1 #2 {{\em Astron. J.\/} {\bf #1}, #2}
\def \ANNREV #1 #2 {{\em Ann. Rev. Astron. Astrophys.\/} {\bf #1}, #2}
\def \APJ #1 #2 {{\em Astrophys. J.\/} {\bf #1}, #2}
\def \APJL #1 #2 {{\em Astrophys.. J. Lett.\/} {\bf #1}, L#2}
\def \APJS #1 #2 {{\em Astrophys. J. Suppl.\/} {\bf #1}, #2}
\def \APSS #1 #2 {{\em Astrophys. Space Sci.\/} {\bf #1}, #2}
\def \ASR #1 #2 {{\em Adv. Space Res.\/} {\bf #1}, #2}
\def \BAIC #1 #2 {{\em Bull. Astron. Inst. Czechosl.\/} {\bf #1}, #2}
\def \JSQRT #1 #2 {{\em J. Quant. Spectrosc. Radiat. Transfer\/} {\bf #1}, #2}
\def \MN #1 #2 {{\em Mon. Not. R. Astr. Soc.\/} {\bf #1}, #2}
\def \MEM #1 #2 {{\em Mem. R. Astr. Soc.\/} {\bf #1}, #2}
\def \PLR #1 #2 {{\em Phys. Lett. Rev.\/} {\bf #1}, #2}
\def \PASJ #1 #2 {{\em Publ. Astron. Soc. Japan\/} {\bf #1}, #2}
\def \PASP #1 #2 {{\em Publ. Astr. Soc. Pacific\/} {\bf #1}, #2}
\def \NAT #1 #2 {{\em Nature\/} {\bf #1}, #2}
\documentstyle{article}
\begin{document}
\title{Magnetic fields of neutron stars: very low and very high}
\author{G.S.Bisnovatyi-Kogan \thanks
{Institute of Space Research, Moscow, Russia; Email:
{\it gkogan@mx.iki.rssi.ru}}}
\date{}
\maketitle
\bigskip
\begin{abstract}
Estimations of magnetic fields of neutron
stars, observed as radio and X-ray pulsars, are discussed.
It is shown, that theoretical and observational values for different
types of radiopulsars are in good correspondence. Magnetic fields of
X-ray pulsars are estimated from the cyclotron line energy. In the case of
Her X-1 this estimation exceeds considerably the value of its magnetic field
obtained from long term observational data related to the beam structure
evolution. Another interpretation of the cyclotron feature, based on the
relativistic dipole radiation mechanism, could remove this discrepancy.
Observational data about soft gamma repeators and their interpretation as
magnetars are critically analyzed.
\end{abstract}

\section{Introduction}
First theoretical estimations of neutron star magnetic fields have been
obtained from the condition of the magnetic flux conservation during
contraction of a normal star to a neutron star (Ginzburg, 1964):

\begin{equation}
\label{ref1}
B_{ns}=B_s\left(\frac{R_s}{R_{ns}}\right)^2.
\end{equation}
For main sequence
stellar magnetic field $B_s=10 \div 100$ Gs, and stellar radius
$R_s=(3 \div 10)R_{\odot}\approx (2 \div 7)10^{11}$ cm, we get
$B_{ns}\approx 4\times 10^{11} \div 5\times 10^{13}$ Gs, for a neutron star
radius $R_{ns}=10^6$ cm. As shown below, this simple estimation occurs to
be in a good correspondence with most observational data.

\section{Radio pulsars}
It is now commonly accepted that radio pulsars are rotating
neutron stars with inclined magnetic axes. For rotating with angular
velocity $\Omega$ dipole,
the rotational energy losses
(Landau \& Lifshitz, 1962; Pacini, 1963) are determined
as $\dot E= A B^2 \Omega^4$. The rotational energy is
$E=\frac{1}{2} I\Omega^2$,
$I$ is an inertia momentum of the neutron star. By measuring of
$P=2\pi/\Omega$ and $\dot P$, we obtain the observational estimation
of the neutron star magnetic field as

\begin{equation}
\label{ref2}
B^2=\frac{IP\dot P}{4\pi^2 A}.
\end{equation}
It was shown by Goldreich and Julian (1969) that energy losses by
a pulsar relativistic wind are important also when the magnetic and
rotational axes coincide, and one may use (\ref{ref2}) with
$A=\frac{R_{ns}^6}{6c^3}$, and $B_{ns}$ corresponding to the magnetic pole,
at any inclination angle. Magnetic fields of radio pulsars estimated
using (\ref{ref2}) with observational values of $P$ and $\dot P$ lay in a
wide region between $10^8$ and $10^{13}$ Gs (Lyne, Smith, 1997).
There are two distinctly different groups: single radiopulsars with
periods exceeding 0.033 sec, and magnetic fields between $10^{11}$ and
$10^{13}$ Gs, and recycled pulsars (RP), present or former members of
close binary systems with millisecond periods and low magnetic fields between
$10^8$ and $10^{10}$ Gs. Low magnetic field of recycled pulsars is
probably a result of its damping during preceding
accretion stage (Bisnovatyi-Kogan, Komberg,
1974, 1976).

\section{X-ray binaries}
There are several ways to estimate observationally magnetic
field of an X-ray pulsar. During accretion matter is stopped by the
magnetic field at the alfvenic surface, where gaseous and magnetic pressures
are in a balance. At stationary state Keplerian angular velocity of
the accretion disk at the alfvenic surface is equal to the stellar
angular velocity (Pringle, Rees, 1972) $\Omega_K=\Omega_s=\Omega_A$.
Otherwise neutron star would be accelerated due to
absorption of matter with large angular momentum, or decelerated due
to throwing away matter with additional angular momentum
(Ilarionov, Sunyaev, 1975). X-ray pulsars may have spin-up and spin-down
stages (Bildsten et al., 1997), but most of them show average
spin-up, what indicate to
their angular velocity being less than the equilibrium one.
This is observed in the best studied X-ray
pulsar Hex X-1 (Sheffer et al.,
1992; Deeter et al., 1998). Analysis of spin-up/down phenomena in the
X-ray pulsars indicate to important role of the mass loss (Lovelace et al.,
1995), and to stochastic origin of spin-up/down transitions (Lovelace et al.,
1999).

For a given luminosity $L_{36}=L/(10^{36}$ergs/s) and dipole magnetic
field at stellar equator $B_{12}=B/(10^{12}$Gs) we get the following
value of the equilibrium period (Lipunov, 1992), for the neutron
star mass $M_s=1.4 M_{\odot}$

\begin{equation}
\label{ref3}
P_{eq}\approx 2.6\,B_{12}^{6/7}\,L_{36}^{3/7} {\rm s}.
\end{equation}
For Her X-1 parameters $L_{36}=10,\,\,P=1.24$ s, we get a magnetic
field corresponding to the equilibrium rotation $B_{12}^{eq}=0.9$.
Taking into account the average spin-up of the pulsar in Her X-1, we
may consider this value as an upper limit of its magnetic field.
Even more rough estimations of the magnetic field in X-ray pulsars
follow from the average spin-up rate under condition $\dot J_{rot}=
\dot M\,\Omega_A$, or restrictions on the polar magnetic field value
following from the observed beam  structure and condition of local
luminosity not exceeding the critical Eddington one. These conditions
lead to smaller values of the magnetic field of Her X-1 on the level
$10^9\, -\, 10^{10}$ Gs (Bisnovatyi-Kogan and Komberg, 1974, 1975:
Bisnovatyi-Kogan, 1974).

Observations of low mass X-ray binaries (LMXB)
indicate to very low values of their magnetic fields due to absence
of X-ray pulsar phenomena. Modulation of X-ray flux permitted to
reveal the rotational period of the neutron star in the LMXB
SAX J1808.4-3658 corresponding to the frequency 401 Hz, due to
RXTE observations (see review of van der Klis, 1998). This observations
fill a gap and form a long-waiting link between LBXB and recycled
millisecond pulsars
(Ruderman, Shaham, 1983), as neutron stars with very low
magnetic field (up to $10^8$ Gs).

The most reliable estimation of the magnetic field of the X-pulsar Her X-1
comes from detailed observations of the beam variation in this pulsar on
different stages, made on the satellites ASTRON (Sheffer et al., 1992),
and GINGA (Deeter et al., 1998). This pulsar, in addition to 1.24 s
period of the neutron star rotation, is in a binary system with an orbital
period 1.7 days, and shows a 35 day cycle, where during only 12 days its
luminosity is high. During other 23 days its X-ray luminosity strongly
decreases, but small changes in the optical luminosity and remaining
strong reflection effect indicate, that the X-ray luminosity
remains almost the same during all 35 day cycle. Visible
decrease of the X-ray flux is due to an occultation phenomena.
The model which explains satisfactory the phenomenon of the 35 day cycle is
based on the precession of the accretion disc with the 35 day period, and
occultation of the X-ray beam during 23 days. Analysis of the beam
structure during high and low X-ray states lead to the conclusion, that
during the low state we observe not the direct X-ray flux from the
neutron star, but the flux, reflected from the inner edge of the
accretion disc. This conclusion is based on the $180\deg$ phase shift
between the X-ray beams in high and low states (Sheffer et al., 1992;
Deeter et al., 1998). In order to observe the X-ray flux reflected from
the inner edge, situated near the Alfven radius of the accretion disc,
it cannot be very far away from the neutron star. The
estimations give the upper limit to the ratio of the Alfven and stellar
radiuses

\begin{equation}
\label{ref4}
\frac{r_A}{r_s} \leq 20.
\end{equation}
The schematic picture of the accretion disc and its inner edge orientation
around the neutron star at different stages of the 35 day cycle
is shown in Fig.1, taken from Sheffer et al. (1992). As was indicated
above, the value of the Alfven radius is determined by the neutron star mass
$(M_s=1.4 M_s)$, mass flux $\dot M=3 \times 10^16$ g/s, corresponding to
the luminosity $L=10^{37}$ ergs/s, and the value of the magnetic field.
Taking dipole radial dependence of the magnetic field $B=B_s(r_s/r)^3$,
and neutron star radius $r_s=10^6$ cm, we obtain the ratio
in the form $r_A/r_s\approx 300 B_{12}^{4/7}$. To have this ratio not
exceeding 20 we get an inequality $B \leq 3\times 10^{10}$ Gs.

Tr\"umper et al. (1978) had found a feature in the spectrum of Her X-1
at energies between 50 and 60 keV. Interpreting it as a cyclotron line
feature according to $E_X=\frac{\hbar e B}{m_e c}$ leads to the
value of the magnetic field
$B_{cycl}=(5-6)\times 10^{12}$ Gs, what is much higher
than any other above mentioned estimations. Spectral features had been
observed also in other X-ray sources. Recent observations on RXTE
(Heindl et al., 1999), and Beppo-SAX (Santangelo et al., 1999)
of the pulsating transient 4U 0115+63 had shown a presence of
3 and 4 cyclotron harmonics features, corresponding to the
magnetic field strength of $1.3\times 10^{12}$ Gs.
A comparison of the shapes of the beam in cyclotron harmonics may
be used for testing the nature of these features.
Cyclotron features had been observed in several X-ray sources
(Mihara et al., 1997), and
they always had corresponded to large values of magnetic fields
$B_{cycl}>10^{12}$ Gs. Such situation was not in good accordance with
a well established observational fact, that all recycled pulsars,
going through a stage of an X-ray source, have much smaller magnetic
fields, usually not exceeding $10^{10}$ Gs. Particularly, for the Her X-1
the value of its magnetic field, following from the cyclotron interpretation
of the spectral feature, was in contradiction with all other observations,
including the most reliable, based on the beam shape variability
during 35 day cycle.

\section{Relativistic dipole interpretation of
        the spectral feature in Her X-1}

It seems likely that this conflict is created by using the
non-relativistic formula connecting cyclotron frequency with a value of
the magnetic field. According to Bisnovatyi-Kogan and Fridman (1969),
ultrarelativistic electrons with a temperature $\sim 10^{11}K$,
may be formed in the non-collisional shock during accretion,
emitting a relativistic dipole line.
The mean energy of this line is broadened and shifted
relativistically, in comparison with the cyclotron line,
by a factor of $\gamma \simeq \frac{kT}{mc^2}$.
The spectrum profile of the relativistic dipole line is calculated
by Baushev and Bisnovatyi-Kogan (1999) for
various electron distributions, where the model of the hot spot
 of Her X-1 is considered, and it is shown that the overall observed
X-ray spectrum (from 0.2 to 120 KeV) can arise under the fields near
$5\cdot 10^{10}$~Gs which are well below $B_{cycl}$, and are not
in the conflict with other observations.

According to Gnedin and Sunyaev (1973), and Bisnovatyi-Kogan (1973),
in the magnetic field near the pulsar the cross
component of a momentum of electrons is emitting rapidly,
while the parallel velocity remains constant.
Hence the momentum distribution of the electrons is anisotropic
 $p_\perp^2\ll p_\parallel^2$, with
$p_\perp\ll mc$, $p_\parallel\gg mc$.
Assume for simplicity that the transverse electron
distribution is two-dimensional Maxwellian
$
  dn=\frac{N}{T_1}{\exp \left( -\frac{m u^{2}}{2T_1} \right)}\,
  d\frac{m u^{2}}{2}
$, $T_1 \ll mc^2$.

In the relativistic dipole regime of
radiation the electron is non-relativistic in
the coordinate system, connected with the Larmor circle. In the
laboratory system, where the electron is moving with a velocity $V$,
the angle between the magnetic field and electron momentum
vectors is smaller than the angle of the emitting beam. In this conditions
we may consider that all radiation is emitted along the magnetic field with
an intensity $J(0)$, after integration over $du$,
at frequency $\omega_{md}$  as (Baushev and Bisnovatyi-Kogan, 1999)

\begin{equation}
\label{ref5}
J(0)=\frac{2 N e^4 B^2 T_1}{\pi c^5 m^3 (1-\frac{V}{c})},\qquad
\omega_{md}=\omega_{cycl} \sqrt \frac{1+\frac{V}{c}}{1-\frac{V}{c}}
\approx 2 \omega_{cycl}\frac{E_\parallel}{m_e c^2},
\end{equation}
where
$\omega_{cycl}=\frac{eB}{m c}$.
That gives
$1-\frac{V}{c}=\frac{2 \omega_{cycl}^2}{\omega^2}$.
Let us consider the parallel momentum distribution of the electrons as:
$dn=f(p_\parallel)\,dp_\parallel$.
Substituting of $dn$ for $N$ and using
$p_\parallel=\frac{mc}{2}\frac{\omega}{\omega_{cycl}}\,$,
we obtain for the spectral density

\begin{equation}
\label{ref6}
J_\omega=\frac{e^2 T_1}{2 \pi c^2 \omega_{cycl}} \omega^2
 f \left( \frac{m c }{2}\frac{\omega}{\omega_{cycl}}\right )\,d\omega.
\end{equation}
Let us consider two cases. The first is a relativistic Maxwell
$f=\frac{n_0 c}{T_2} \exp\left( -\frac{p_\parallel c}{T_2}\right )$,
where $n_0$ is a number of emitting electrons. The spectrum is

\begin{equation}
\label{ref7}
J_\omega=\frac{n_0 e^2}{2 \pi c \omega_{cycl}}\frac{T_1}{T_2} \omega^2
 \exp\left( -\frac{m c^2 \omega}{2 \omega_{cycl} T_2}\right )
\,d\omega.
\end{equation}
It has a single maximum at
$\frac{\omega}{\omega_{cycl}}=\frac{4 T_2}{m c^2}$. In the second case\\
$f=\frac{n_0 }{\sqrt {\pi} \sigma}
\exp\left[ -\frac{{(p_\parallel-a)}^2}{\sigma^2}\right ]$.
The spectrum of radiation is

\begin{equation}
\label{ref8}
J_\omega=\frac{n_0 e^2 T_1}{2 \pi^{3/2} c^2 \omega_{cycl}\sqrt{\sigma}}
\omega^2
 \exp\left( -\frac{(\frac{m c}{2}\frac{\omega}{\omega_{cycl} }-a)^2}
 {\sigma^2}\right )
\,d\omega.
\end{equation}
When $\sigma \ll a$ this spectrum has a single maximum at
$\omega \simeq \frac{2 a}{m c} \omega_{cycl}$.
Baushev and Bisnovatyi-Kogan (1999)
had approximated experimental spectra taken from Mihara et al. (1990),
and McCray et al. (1982). The last spectrum (solid line)
and its fitting (dashed line) are shown in Fig.2.
It was taken in accordance with  Bisnovatyi-Kogan and Fridman (1969),
$a=7 \cdot 10^{-4}$~${\rm\frac{eV \cdot s }{cm}}$, corresponding to
average electron energy $E_{\parallel}=ac \approx 20$ MeV,
and the best fit for the line shape was obtained at
the magnetic field strength $B=4\cdot 10^{10}$~Gs.
In this model the beam of the "cyclotron" feature is determined by the
number distribution of the emitting relativistic electrons,
moving predominantly
along the magnetic field, over the polar cap.

In order to obtain the whole experimental spectrum of the Her X-1
the following model of the hot spot (Fig.3) was considered
by Baushev and Bisnovatyi-Kogan (1999).
A collisionless shock wave is generated in the accretion flow
near the surface on the magnetic pole of a neutron star. In it`s front the
ultrarelativistic electrons are generated.
Under the shock
there is a hot turbulent zone with a temperature $T_e$,
and optical depth $\tau_e$, situated over a heated spot
with a smaller temperature on the surface
of the neutron star.

The whole X-ray spectrum of pulsar Her X-1 from McCray et al. (1982)
is represented in Fig.3 by the solid line.
There are three main regions in it:
a quasi-Planckian spectrum between 0,3 and 0,6 KeV, that is generated
(re-radiated) near the magnetosphere of the X-ray pulsar;
power-law spectrum $(0.6\div
20)$~KeV with a rapid decrease at 20 KeV, and the "cyclotron" feature.
The power-law spectrum is a result of comptonization in the corona
of a black-body spectrum emitted by the stellar surface.
The comptonized spectrum has been calculated according to
Sunyaev and Titarchuk (1980).
Setting the neutron star radius equal to 10 km, distance to the
X-ray pulsar 6 Kps, hot spot area $S=2 \cdot 10^{12}$~cm$^2$,
the best fit was found
at $T_s=1$~KeV, $T_e=8$~KeV, $\tau_e=14$, which is
represented in Fig.3 by the dashed line.

The observations of the variability of the "cyclotron" lines
are reported by Mihara et al. (1997). Ginga detected the
changes of the cyclotron energies
in 4 pulsars. The change is as much as 40 \% in the case of 4U 0115+63.
Larger luminosity of the source corresponds to smaller average energy of the
cyclotron feature. These changes might be easily explained in our model.
The velocity of the accretion flow decreases with increasing of the
pulsar
luminosity because locally the luminosity is close to the Eddington limit.
As a result the shock wave intensity drops as well as the energy of the
ultrarelativistic electrons in it`s front, leading to
decrease of the relativistic dipole line energy.

\section{Magnetars}

Among more than 2000 cosmic gamma ray bursts (GRB) 4 recurrent sources
had been discovered, and were related to a separate class of GRB, called
soft gamma repeators (SGR). Besides observations of short, soft, faint
recurrent bursts, three of them had given giant bursts, most powerful among
all GRB. All four SGR are situated close or inside SNR, three of them show
long periodic pulsations. These properties had separated SGR, situated in our
or nearby galaxies, in a quite special class, very different from other
GRB, which are believed to have a cosmologilal origin at red shifts $z \sim 1$.
One SGR 1627-41 had been discovered by BATSE (Woods et al., 1999a) and
three other had been discovered in KONUS experiment in 1979 (Mazets et al.,
1979, 1981; Golenetskii et al., 1984). Three of SGR show regular pulsations,
and for two of them $\dot P$ had been estimated, indicating to very
high values of the magnetic fields, up to $10^{15}$ Gs, and small age of
these objects.
SGR have the following properties
(Feroci et al., 1999; Hurley et al., 1999a-e;
Kouveliotou et al., 1998, 1999; Mazets et al., 1979, 1999a-c; Murakami
et al., 1999; Woods et al., 1999b)

\quad 1. SGR0526-66 (Mazets et al., 1979, 1999c)

It was discovered due to a giant burst of 5 March 1979, projected to the
edge of the SNR N49 in LMC. Accepting the distance 55 kpc to LMC, the peak
luminosity in the region $E_{\gamma}>30$ keV is equal to
$L_p=3.6\times 10^{45}$ ergs/s, the total energy
release in the peak $Q_p=1.6 \times 10^{44}$ ergs, in the subsequent tail
$Q_t=3.6 \times 10^{44}$ ergs. The short recurrent bursts have
peak luminosities in this region
$L_p^{rec}=3\times 10^{41}\,-\, 3 \times 10^{42}$ ergs/s,
and energy release $Q^{rec}=5\times 10^{40}\,-\, 7 \times 10^{42}$ ergs.
The tail was observed about 3 minutes and had regular pulsations with the
period $P\approx 8$ s. There was not a chance to measure $\dot P$ in this
object.

\quad 2. SGR1900+14 (Mazets et al., 1999c,d; Kouveliotou et al., 1999;
  Woods et al., 1999b)

It was discovered first due to its recurrent bursts, the giant burst
was observed 27 August, 1998. The source lies close to the
less than $10^4$ year
old SNR G42.8+0.6, situated at distance $\sim 10$ kpc.
Pulsations had been observed in the giant burst, as well as in the
X-ray emission observed in this source in quiescence by RXTE and ASCA.
$\dot P$ was measured, being strongly variable.
Accepting the
distance 10 kpc, this source had in the region $E_{\gamma}>15$ keV:
$L_p > 3.7\times 10^{44}$ ergs/s,
$Q_p > 6.8\times 10^{43}$ ergs,
$Q_t=5.2 \times 10^{43}$ ergs,
$L_p^{rec}=2\times 10^{40}\,-\, 4\times 10^{41}$ ergs/s,
$Q^{rec}=2\times 10^{39}\,-\, 6\times 10^{41}$ ergs,
$P=5.16$ s,
$\dot P=5 \times 10^{-11}\,-\, 1.5\times 10^{-10}$ s/s.
This source was discovered at frequency 111 MHz as a faint,
$L_r^{max}=50$ mJy, radiopulsar
(Shitov, 1999) with the same $P$ and variable $\dot P$ good corresponding to
X-ray and gamma-ray observations. These values of $P$ and
average $\dot P$, according
to (\ref{ref2}) correspond to the rate of a loss of rotational energy
$\dot E_{rot}=3.5 \times 10^{34}$ ergs/s, and magnetic field
$B=8 \times 10^{14}$ Gs. The age of the pulsar estimated as
$\tau_p=P/2\dot P=700$ years is much less than the estimated age of the
close nearby SNR. Note that the X-ray luminosity of this object
$L_x=2\times 10^{35}\,\, - \,\, 2\times 10^{36}$ ergs/s
is much
higher, than rate of a loss of rotational energy, what means that rotation
cannot be a source of energy in these objects. It was suggested that the
main source of energy comes from a magnetic field annihilation, and such
objects had been called as magnetars (Duncan, Thompson, 1992).

\quad 3. SGR1806-20 (Kouveliotou, 1998; Hurley et al., 1999e)

This source was observed only by recurrent bursts. Connection with
the Galactic radio SNR G10.0-03 was found. The source has a small but
significant displacement from that of the non-thermal core of this SNR.
The distance to SNR is estimated as 14.5 kpc. The X-ray source observed
by ASCA and RXTE in this object shows regular pulsations with a period
$P=7.47$ s, and average $\dot P=8.3\times 10^{-11}$ s/s. As in the previous
case, it leads to the pulsar age $\tau_p \sim 1500$ years, much smaller
that the age of SNR, estimated by $10^4$ years. These values of $P$ and $\dot
P$ correspond to $B=8\times 10^{14}$ Gs. $\dot P$ is not constant, uniform set
of observations by RXTE gave much smaller and less definite value
$\dot P=2.8(1.4)\times 10^{-11}$ s/s, the value in brackets gives 1$\sigma$
error. The peak luminosity in the burst reaches
$L_p^{rec}\sim 10^{41}$ ergs/s in
the region 25-60 keV, the X-ray luminosity in 2-10 keV band is
$L_x\approx 2\times 10^{35}$ ergs/s is also much higher than the rate
of the loss of rotational energy (for average $\dot P$) $\dot E_{rot}
\approx 10^{33}$ ergs/s.

\quad 4. SRG1627-41 (Mazets et al., 1999a; Woods et al., 1999a)

Here the giant burst was observed 18 June 1998, in addition to numerous
soft recurrent bursts. Its position coincides with the SNR G337.0-0.1,
assuming 5.8 kpc distance. Some
evidences was obtained for a possible periodicity of 6.7 s, but giant burst
does not show any periodic signal (Mazets et al., 1999a), contrary to
two other giant burst in SGR. The following characteristics
had been observed with a time resolution 2 ms at photon energy
$E_{\gamma}> 15$ keV:
 $L_p \sim 8\times 10^{43}$ ergs/s,
$Q_p \sim 3\times 10^{42}$ ergs,
no tail of the giant burst had been observed.
$L_p^{rec}=4\times 10^{40}\,-\, 4\times 10^{41}$ ergs/s,
$Q^{rec}=10^{39}\,-\, 3\times 10^{40}$ ergs. Periodicity in this source is
not certain, so there is no $\dot P$.

To measure $\dot P$ the peaks of the beam are compared during long
period of time. Both SRG with measured $\dot P$ have highly variable beam
shapes, what implies systematic errors in the result. Another source of
systematic error comes from the barycenter correction of the arriving
signal in
the source with an essential error in  angular coordinates. This effect is
especially significant for determination of $\ddot P$ (Bisnovatyi-Kogan,
Postnov, 1993), but when observational shifts are short the error in the
coordinates could not be extracted easily. Earth motion around the Sun, as
well as the satellite motion around the Earth may influence the results.
Nevertheless, independent measurements of $P$ and $\dot P$ in such
different spectral bands as radio and X-rays gave similar results for
SGR1900+14.

The physical connection between SGR and related SNR is not perfectly
established: SNR ages are much larger, than ages of SNR estimated
by $P$ and $\dot P$ measurements, and all four SGR are situated at the
very edge of the corresponding SNR, or well outside them. Using the pulsar
age estimation we come to conclusion of a very high speed of the neutron star
at several thousands km/s, exceeding strongly all measured speeds of
radiopulsars.
Physical properties of the pulsars observed in SGR are very unusual.
If the connection with SNR is real, and the distances to SGR are the same
then the pulsar luminosity
during giant bursts is much larger than the
critical Eddington luminosity. As follows from simple physical reasons,
when the radiation force is much larger than the force of gravity, the matter
would be expelled at large speed, forming strong outflow and dense envelope
around the neutron star which could screen pulsations. No
outflowing envelope around SGR have been found.
The  large difference between $\dot E_{rot}$ and average $L_x+L_{\gamma}$
luminosity needs to suggest a source of energy, much larger than the one
coming from the rotational losses. It is suggested that in magnetars the
energy comes from the magnetic field annihilation
(Duncan, Thompson, 1992). It is rather surprising to observe the
field annihilation without formation of relativistic
particles, where considerable part of the released energy should go.
Radio emitting nebula should be formed around the SGR, but
had not been found.  So, it is not possible to exclude that there is no
physical connection between SNR and SGR, that SGR are much closer objects,
their pulsar luminosity is less than the Eddington one, and their
magnetic fields are not so extremely high.

There is a striking similarity between SGR and special class of
X-ray pulsars called anomalous X-ray pulsars (AXP). Both have periods
in the interval 6 - 12 seconds, binarity was not found, spin-down of the
pulsar corresponding to small age $\sim 1000$ years, observed irregularities
in $\dot P$ (see e.g. Melatos, 1999). The main difference is an absence of
any visible bursts in AXP, characteristic for SGR. Suggestions had been
made about their common
origin as magnetars (Kouveliotou et al., 1998, Melatos, 1999). Other
models of AXP had been discussed (see e.g. Mereghetti et al., 1998). We
may expect that establishing of the nature of AXP would help strongly
for the determination of the nature of SGR.

\section{Conclusions}

\begin{enumerate}
\item Magnetic fields of radiopulsars are in good correspondence with
theoretical estimations.
\item  RP and LMXB have small magnetic fields, which very probably
had been decreased by damping or screening during accretion stage.
\item  Contradiction between high $B_{cycl}$ and other
observational estimations of $B$ in the LMXB Her X-1 may be removed in
the model of relativistic dipole mechanism of the formation of a hard
spectral feature by strongly anisortopic relativistic electrons, leading to
conventional value of $B\approx 5\times 10^{10}$ Gs.
\item  Very high magnetic fields in magnetar model of SGR needs farther
confirmation and investigation.
\end{enumerate}

\section{Questioning, Answering, and Checking magnetars}

\begin{enumerate}
\item {\bf Q}. Is it possible to observe an influence of a giant burst on the
related SNR ?\\
\qquad {\bf A}. May be. \\
\qquad {\bf C}. Radio observations of SNR, search for changes,
wisps like in the Crab nebula.
\item {\bf Q}.What could we see if SGR would be 10 times farther ?\\
\qquad {\bf A}. Ordinary short GRB.
\item {\bf Q}. Why observed short GRB has larger hardness among GRB,
opposite to SGR (Cline et al., 1999)?\\
\qquad {\bf A}. We could see only the main peak which is rather hard.
\item {\bf Q}. How many ``SGR'' we should have seen in neighboring galaxies
(Andromeda) by their giant busts?\\
\qquad {\bf A}. $\sim 30$ in Andromeda.\\
\qquad {\bf C}. Better statistical estimation and search in BATSE/KONUS data.
\item {\bf Q}. Are short GRB identical with distant SGR?\\
\qquad {\bf A}. May be yes.
\item {\bf Q}. How many SGR remnants should be in the Galaxy if SGR-SNR
connection is real?\\
\qquad {\bf A}. About $10^8$ neutron stars.
\item {\bf Q}. Is it possible to provide $\gamma$-radiation by magnetic
field annihilation without appearance of large number of ultra-relativistic
particles and strong nonthermal emission in other spectral bands?\\
\qquad {\bf A}. Solar flashes produced by field annihilation are
characterized by very broad spectrum, from radio to gamma.
\item {\bf Q}. Is it possible to use pulsar-like formula for age estimation
of SGR and its magnetic field from $P$ and $\dot P$ measurements, when
$L_{tot}\gg \dot E_{rot}$?\\
\qquad {\bf A}. Probably not (Marsden et al., 1999).
\end{enumerate}

{\bf Acknowledgements}
I am very grateful to Franco Giovannelli and other organizers of Vulcano99
workshop for their kind hospitality and support.

\vfill\eject

{\bf Figure captions}

\medskip
{\bf Fig.1} Configuration of the inner edge of the disk and the neutron star;
neutron star and the disk in the "high-on" state (left top box),
and in the "low-on" state (right bottom box).

\medskip
{\bf Fig.2} Comparison of the observational and computational X-ray spectra
of Her X-1. The solid curve is the observational results taken from
McCray et al. (1982),
the dot curve is the approximation with
$T_s=0.9$~KeV, $T_e=8$~KeV, $\tau_e=14$,
$a=7 \cdot 10^{-4}$~${\rm\frac{eV \cdot s }{cm}} $,
$\sigma=10^{-4}$~${\rm\frac{eV \cdot s }{cm}} $, $B=4 \times 10^{10}$ Gs.

\medskip
{\bf Fig.3} Schematic structure of the accretion column near the
magnetic pole of the neutron star (top), and its radiation spectrum (bottom).

\end{document}